# Quantum-dot Cellular Automata (QCA): A Survey

[1]Usha Mehta, [2]Vaishali Dhare

[1] *Professor, Institute of Technology, Nirma University, Ahmedabad, India*
[2] *Ph.D. Scholar, Institute of Technology, Nirma University, Ahmedabad, India*
[1]usha.mehta@nirmauni.ac.in   [2]vaishali.dhare@nirmauni.ac.in

*Abstract*— *In the near future the era of "Beyond CMOS" will start as the scaling of the current CMOS technology will reach the fundamental limit. QCA (Quantum-dot Cellular Automata) is the transistor less computation paradigm and viable candidate for "Beyond CMOS" device technology.*

*The complete state of art survey on QCA is presented in this paper. This paper addresses the QCA background, its possible implementation and available simulation and synthesis tools. In depth survey is carried out for the QCA oriented defects and testing. Also, need of development and possible research areas in various sides of QCA are discussed.*

## I. INTRODUCTION

Continued and fast dimensional scaling of CMOS eventually will approach the fundamental limit [1]. Also, Short channel effects, high power dissipation, quantum effects are limiting the further scaling of current CMOS technology devices [2-3]. Emerging device technology can overcome the scaling limitation in the current CMOS technology [1]. Single Electron Transistor (SET) [4], Quantum-dot Cellular Automata (QCA) [5] and Resonant Tunneling Diodes (RTD) [6] are some of the "Beyond CMOS" technologies. Among these evolving nanotechnologies, Quantum-dot Cellular Automata is the most favorable technology [1]. QCA is transistorless computational paradigm which can achieve device density of $10^{12}$ devices/cm$^2$ and operating speed of THz. QCA device paradigm replaces FET based logic and exploit the quantum effects of small size.

Quantum-dot Cellular Automata is a mean of representing binary information on cells, through which no current flows, and achieving device performance by the coupling of those cells [5,7].

This paper presents the state of art survey on QCA basics, implementation, fabrication, tools, defect characterization, fault model and testing. Also the paper addresses the issues in some of the methods and techniques. Further, the paper suggests the possible research area of QCA.

The rest of the paper is organized as follows, Section II describes the QCA background. Section III describes QCA defect analysis, fault model and testing. The paper concludes in Section IV.

## II. QCA BACKGROUND

### A. QCA Cell

Unlike current switching in CMOS technology, QCA encodes the binary information as per the position of individual electrons. QCA is the array of cells in which each cell consists of quantum dots also considered as sites that are positioned at the corners of the square cell. The charge is localized in the dots. Also, the cell consists of two mobile electrons that can tunnel between the dots. Electron tunneling out of the cell is not possible due to the potential barriers between cells. Two free electrons resides at the corners of the cells, always diagonally due to Coulombic repulsion. Cell configurations with four, five and six are available [5,7-8]. The four dot QCA cell with quantum dot's number (site) is shown in Fig. 1 (a). The polarization (P) of the cell is calculated by equation (1) [8].

$$P = \frac{(\rho_1 + \rho_3) - (\rho_2 + \rho_4)}{\rho_1 + \rho_2 + \rho_3 + \rho_4} \quad (1)$$

Where $\rho i$ is expectation value of the number operator on site (dot) for the ground state eigenfunction as given by equation (2). Where i is the quantum dot's number 1, 2, 3, and 4 as depicted in Fig. 1 (a).

$$\rho_i = \langle \psi_0 | \hat{n}_i | \psi_0 \rangle \quad (2)$$

Where $|\psi_0\rangle$ is ground state of the cell and it is given in the equation (3)

$$|\psi_0\rangle = \sum_j \psi_j^0 |\phi_j\rangle \quad (3)$$

Where $|\phi_j\rangle$ is the $j^{th}$ basis vector and $\psi_j^0$ is the coefficient of the basis vector. It is determined by direct diagonalization of the Hamiltonian.

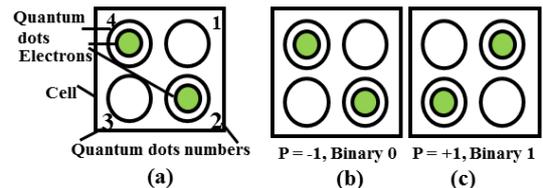

Fig. 1. QCA cell (a) schematic (b) with polarization P = "-1" (c) with polarization P = "+1".

Electrons are located diagonally for which cell polarization is calculated. If the electrons are located as shown in Fig. 1(b) then according to the equation 1, cell polarization P = -1 is encoded as binary 0 (Logic 0). In the same way, considering the electrons location as shown in Fig. 1(c), the cell polarization P = +1 is encoded as binary 1 (Logic 1). Coulombic coupling between cells causes the information flow in the QCA array.

Cell to Cell response function is shown in Fig. 2. [9].

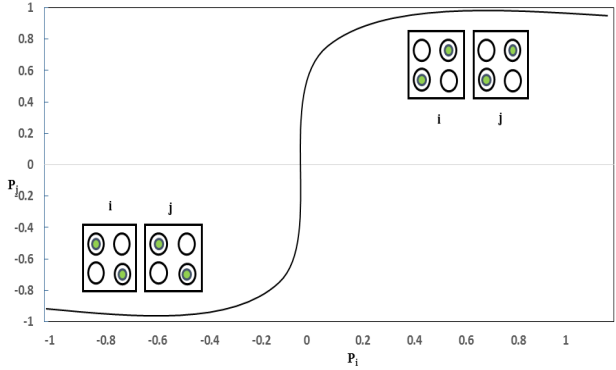

Fig. 2. The cell–cell response [9].

The cell to cell response is computed by solving the two particle Schrodinger equation [8]. In two-cells $i$, $j$ system, polarization of cell $j$ is aligned with its neighbor cell $i$. In this case cell $i$ is considered as a driver. In N-cell system, for single cell $i$, the two state model is calculated by the Hamiltonian given in equation (4).

$$\hat{H}_i = \begin{pmatrix} -\frac{1}{2}\sum_j E_{i,j}^k P_j & -\gamma_i \\ \gamma_i & \frac{1}{2} E_{i,j}^k P_j \end{pmatrix} \quad (4)$$

Where $\gamma_i$ is the tunneling energy, $E_{i,j}^k$ is the Kink energy between cells $i$ and $j$ given by equation (5). $P_j$ is the polarization of cell $j$. The Coulombic interaction between two cells can be described by the kink energy $E_{kink}$. Kink energy is the difference of electrostatic energies of two cells with opposite polarization and same polarization [8].

$$E_{kink}^{i,j} = E_{opposite}^{i,j} - E_{same}^{i,j} \quad (5)$$

$E_{opposite}^{i,j}$ : Energy between cell $i$ & $j$ with opposite polarization.

$E_{same}^{i,j}$ : Energy between cell $i$ & $j$ with same polarization.

The electrostatic energy between two cells is used to find the state energy. The electrostatic energy between cells i and j is given by (6).

$$E^{i,j} = \frac{1}{4\pi\varepsilon_0\varepsilon_r} \sum_{n=1}^{4} \sum_{m=1}^{4} \frac{q_n^i q_m^j}{|r_n^i - r_m^j|} \quad (6)$$

Where $\varepsilon_0$ is the permittivity of free space, $\varepsilon_r$ is the relative permittivity of material, $q_n^i$ is the charge in dot $n$ of cell $i$, $q_m^j$ is the charge in dot m of cell $j$, $r_n^i$ is the position of nth dot in cell $i$, $r_m^j$ is the position of mth dot in cell $j$, thus $|r_n^i - r_m^j|$ is the distance between $n^{th}$ dot in cell $i$ and mth dot in cell $j$.

### B. QCA Implementation

According to the material used to realize QCA cell, the types of QCA are metal island [10, 11, 13-15], molecular [16-20], magnetic [21-23] and semiconductor [9,24]. Among these realizations, Molecular QCA is the most favorable, since it can operate at room temperature. Also, from the fabrication point of view, molecular QCA is the most viable type. Molecular QCA will be able to operate at the speed of THz with ultra low power and extremely high device density [25].

#### 1. Metal QCA

In [10], the Al-AlOx-Al tunnel junctions are fabricated on an oxidized Si substrate by a standard electron beam–lithography and shadow evaporation techniques. This metal QCA cell consists of four aluminum islands as dots, D1 to D4 which are coupled with aluminum oxide tunnel junctions and capacitors. The two dots E1 and E2 are SET electrometers for sensing the output [11]. Ballistic point-contacts, STM method, and SET electrometer can be used to read the output. The cell is shown in Fig. 3. [11, 12]. The metal based QCA latch is implemented and demonstrated in [15] which was operated at the temperature of 70mK.

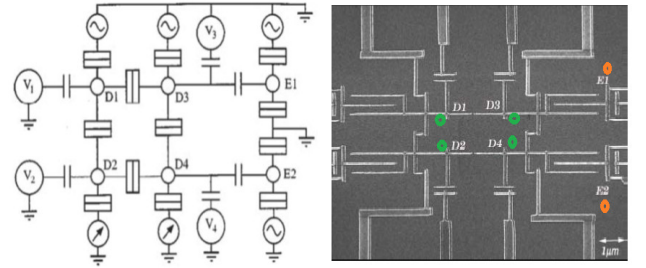

Fig. 3. Metal QCA cell (a) Simplified Schematic Diagram of Four-dot Metal QCA Cell [12] (b) shows the scanning electron micrograph of this QCA cell [13,14].

#### 2. Molecular QCA

Molecular QCA (mQCA) is considered as the promising implementation for QCA circuits. It operates at room temperature with high device density, and high operating speed. Molecular QCA cell is presented in [16-18].

Molecular implementation could also be fabricated with much higher uniformity than those achievable with semiconductor or metal-island QCA implementations. Each molecule acts as a cell in which redox centers acts as dots and tunneling is provided by bridging ligands. Binary information is encoded with charge configurations. Molecular QCA cell is shown in Fig.4. [19, 20].

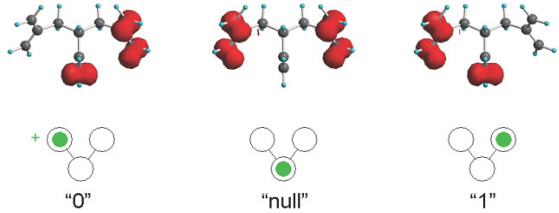

Fig. 4. Three states of six dot molecular QCA [16, 18]

An ab initio quantum chemistry analysis of a molecular QCA, {(η5-C5H5)Fe(η5-C5H4)}4(η4-C4)Co(η5-C5H5)2+ molecule has been carried out in [19].

3. Magnetic

Cowburn and Welland have developed and proposed a magnetic implementation of quantum dot cellular automata (MQCA) [21, 22]. Unlike other types of QCA, it is based on the interaction between magnetic nano-particles. In MQCA, information is propagated through magnetic interactions. It is predicted that magnetic QCA can reach a speed of about 100 MHz. The binary logic representation in magnetic quantum-dot cellular automata is shown in Fig.5. A three-input majority gate in magnetic QCA has been fabricated in [23].

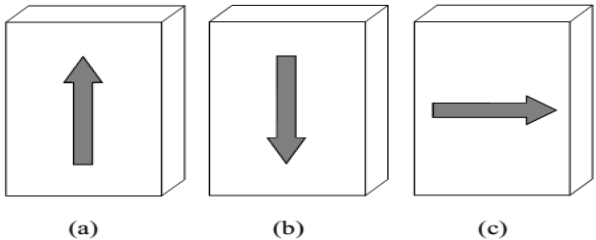

Fig. 5. Binary logic representation in magnetic QCA. (a) Logic '1', (b) Logic '0', and (c) Null state.

4. Semiconductor

Semiconductor quantum dots are nano-structured created from a standard semiconductor material such as InAs/GaAs and GaAs/AlGaAs [9]. In a semiconductor QCA, four semiconductor quantum-dots are placed at four different corners of the substrate. Cell polarization is encoded as charge position, and quantum-dot interaction relies on the electrostatic coupling. The advantage of using semiconductor QCA is that it is based on materials that are well understood and many fabrication techniques have been created to work with them. A possible physics implementation of a QCA cell is shown in Fig 6. The electric field in the substrate is introduced by the top metal gate to deplete electrons in the 2–D electron gas formed at the junction of the dielectric layer and the substrate. Quantum dots are formed at locations where the metal gate is removed to leave an exposed surface. Fabrication using GaAs/AlGaAs heterostructure with a high-mobility two-dimensional electron gas below the surface is demonstrated in [24].

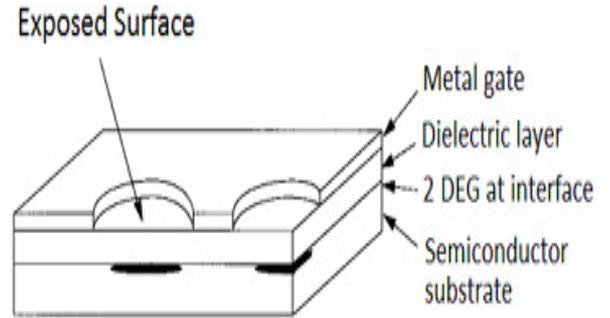

Fig. 6. The physical representation of the semiconductor QCA cell.

C. QCA Clocking

In CMOS technology, the clock is used to control the timing, mostly in sequential circuits. In QCA, Clock provides the switching and power gain to the circuits [26]. The clock signal is given to the each QCA cell in combinational as well as sequential circuits to raise or lower the tunneling barrier between dots. The clock signals are generated using the electric field. It is originated from wires buried below the QCA surface using CMOS wires or $CNT_S$ (Carbon Nano Tube) [27].

In [9], two types of switching method, abrupt and adiabatic are discussed. QCA circuit may enter into metastable state in abrupt switching, hence it is not suitable. Hence, adiabatic switching is used in which the cells are in the non-polarized state at the low barrier and allowed to change the polarization at the high barrier.

The clocking scheme is proposed in [20]. It consists of 4 clock signals or zones and each clock signal consists of four phases namely switch, hold, release and relax as shown in Fig. 7. The frequency of each clock is same with the phase shift of $90^0$ each. One of the clock signals can be considered the reference (phase = 0) and the others are delayed one (phase = $\pi/2$), two (phase = $\pi$) and three (phase = $3\pi/2$) quarters of a period as shown in Fig. 7. Clocking is done by electrostatically switching the cell from a null state, in which cell holds no binary information. In switch phase, the cell state is determined by its neighbors, to a locked state, in which the state is independent of its neighbors. The whole QCA circuit is divided into different clock zones and each clock signal is given to its respective zone. During the information flow, the cells of each clock zone passes through all four phases of that respective clock zone. The information is transfer after these four phases.

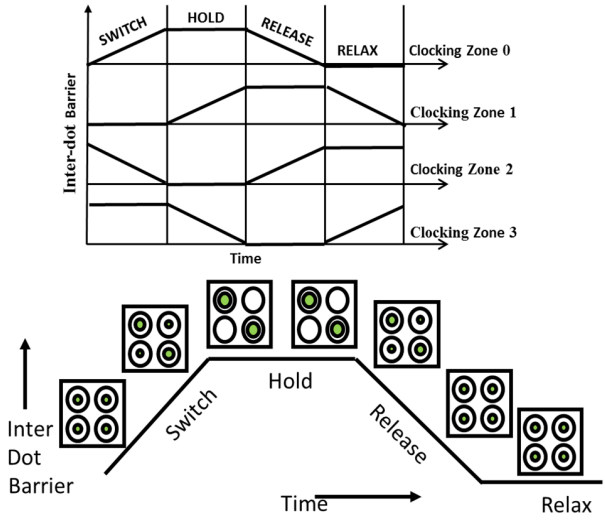

Fig. 7. Clocking scheme.

In the switch phase, the barriers are raised and the cells become polarized according to the polarization of their driver. This is the clock phase during which actual computation occurs. At the end of this clock phase, the barriers are high and it suppress any kind of tunneling. Now the cell polarization is fixed. During the hold phase, the barriers are held at high value and the cell is now acting as an input to the next stage. Next, in the release phase, the barriers are lowered and the cells are allowed to relax to an unpolarized state. In the relaxed phase, the cell barriers remain lowered, keeping the cells in an unpolarized, neutral state. After this fourth phase, the subsystem will return to the first clock phase and repolarize. For reliable kink-free computation the number of cells allowed in one clock zone must be $<= e^{\frac{E_k}{K_B T}}$, Where $E_k$ is kink energy, $K_B$ is Boltzmann constant and T is the operating temperature in degree Kelvin. Two dimensional (2D) clocking scheme is proposed in [28, 29]. It is based on the parallel execution and processing in clocking zones within a different timing framework. Ripple Clock Schemes for QCA Circuits is proposed in [30]. Bennett clocking of QCA has been described in [31].

*Assigning four clock signal causes delay in the final output as per the number of clocks required for the particular circuit. So large circuit may have more delay. Clock zone assignment is the critical part during QCA circuit layout implementation and simulation. Hence, development of concrete and novel clocking scheme is required. Research in advanced clocking scheme can be carried out.*

### D. QCA Basic Elements

The basic building blocks of QCA are Majority voter (MV), inverter and binary wire [32]. Also, $45^0$ inverter chain, coplanar wire crossing and multilayer crossover are proposed in [32].

The MV is the 3-input basic primitives which consists of 5 cell configuration as shown in Fig. 8 (a). The cells A, B, C are input cells, the middle cell is the device cell which has the polarization of the majority of the inputs and right hand side cell is the output cell which has same polarization as device cell.

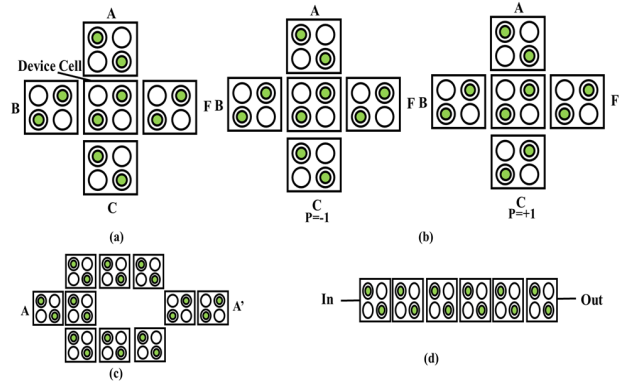

Fig. 8. QCA Elements (a) Majority Voter (b) MV as AND and OR gates (c) inverter (d) binary wire

The output of the MV, as indicated by the name is determined by the majority of its three inputs. MV implements the Boolean function F = AB + BC + AC. Where F is the output and A, B and C are the inputs.
For example, if two of the inputs are low, the output is low. If two of the inputs are high, then output is high. Here high refers to the polarization state P = +1, and low refers to the polarization state P = -1. The truth table of a majority gate covering all possible combination of inputs and corresponding output is shown in Table 1. 2-input AND and OR logic implementations are possible by keeping one of the inputs of MV to fixed polarization P = -1 and P = +1 respectively as shown in Fig. 8(b).

Table 1 Truth table of majority gate

| A | B | C | F |
|---|---|---|---|
| 0 | 0 | 0 | 0 |
| 0 | 0 | 1 | 0 |
| 0 | 1 | 0 | 0 |
| 0 | 1 | 1 | 1 |
| 1 | 0 | 0 | 0 |
| 1 | 0 | 1 | 1 |
| 1 | 1 | 0 | 1 |
| 1 | 1 | 1 | 1 |

QCA inverter is shown in Fig. 8(c), is another fundamental logic gate in QCA with one input and one output. It takes the input logic and produces its complement logic on output. The truth table of it is given in Table 2. If the input is logic high, the output will be low. If the input is logic low, the output will be high. Many implementations are possible for inverter but the inverter shown in Fig. 8(c) is considered as robust as inversion take place in two paths.

Table 2 Truth table for QCA inverter

| In | Out |
|----|-----|
| 0  | 1   |
| 1  | 0   |

The QCA binary wire shown in Fig. 8(d) is used to transfer information from one part of the circuit to another. In a QCA circuit, a wire not only helps in information transfer, it actually can performs some computational operation on the information to be transferred.

### E. Fabrication

The key issue in the QCA fabrication is the expensive nanoscale lithography. Self-assembly method is the alternative to it. The possibility of molecular manipulation through the use of Scanning Acoustic Microscopy (SAMs) and supramolecular Chemistry using self-assembly is presented in [33]. Lithographic resolution of 5 nm using a cold-development technique is demonstrated in [34]. In [34] high-resolution Electron Beam Lithography (EBL) and molecular lift-off is applied to pattern Creutz–Taube molecules on the scale of a few nanometers for QCA. Molecular QCA array by electric field from the FeIII-RuII configuration to the FeII-RuIII configuration is demonstrated in [35]. Two ferrocene and two ferrocenium moieties as a component for charge-coupled QCA circuits as a assembly of a symmetric square Cell is shown in [36]. Ions and the associated counter ions can disrupt the correct flow of information in molecular QCA. Self-doping mechanism which incorporates the counterion covalently into the structure of a neutral molecular cell is presented in [37]. All discussed fabrication of QCA is done as experimental point of view only.

### F. Tools

The research has been done in QCA simulation and synthesis tools. The available tools are MAQUINAS [38], QBert [39], QCA-LG [40], and QCADesigner [41]. MAQUINAS and QCADesigner are rely on a simulation engine that solves the Schrödinger equation for modeling physical interactions within the considered set of cells in the circuit. MAQUINAS uses adiabatic switching and time independent Schrodinger equation is solved to calculate polarization state of each cell. Fast simulator for QCA digital logic, QBART is developed in [39]. Automatic layout generation tool QCA-LG is developed in [40]. QCADesigner [41] is the most widely used QCA layout simulator. It uses two simulation engines namely bistable approximation and coherence vector. All versions of QCADesigner are available on [42]. The library of basic QCA elements in Hardware Description Language, Verilog (HDLQ) is proposed in [43]. This library also incorporated with fault injection and bidirectionality. In [44] a SPICE macro model for QCA has been proposed and experimentally verified. Polarization error and power estimation tool for QCA, QCAPro is proposed in [45]. CAD tool, HDLM for magnetic QCA, based on Verilog HDL is proposed in [46]. HDL models for a magnetic QCA cell and building blocks are proposed which ensure magnetization, clocking, and signal propagation.

Synthesis is the significant process in any digital design flow. Synthesis methods for digital logics into QCA basic primitives are proposed in [47-51]. Zhang et al [48] have proposed AND/OR-based logic synthesis for QCA combinational circuit. This method reduces the number of MVs to compute three variable Boolean functions by simplifying the conversion of Sum-of-Product (SOP) expressions into QCA majority logic. Zhang et al [48] have proposed the logic synthesis tool MALS for MV-based logic. In [48], each decomposed subcircuits using maximum four MVs are implemented. In [49] efficient decomposition scheme is introduced which removes the redundancies produced in the process of converting a decomposed network into a majority network. Minimal majority gate mapping with a Majority expression Look-Up Table (MLUT) based algorithm is developed in [50].

Above discussed methods for logic synthesis are limited for small Boolean functions and so fundamental algebra for n-majority is developed in [51]. Recently, heuristic based majority/minority logic synthesis methodology is proposed in [52]. More research is required for the development of synthesis methods and tools.

*Research in QCA CAD tool development needs much attention and can be carried out. As of now no commercial simulation and synthesis tool is available.*

### III. DEFECT, FAULT, FAULT MODELS AND TESTING OF QCA DEVICES AND CIRCUITS

#### A. Defects

QCA devices are prone to defects due to the nanoscale. The survey of defect characterization has been carried out and presented in [53]. QCA defect classification is shown in Fig. 9. Defects during molecular QCA manufacturing can occur in two phases namely synthesis and deposition. In the synthesis phase, the individual cells (molecule) are manufactured and in the deposition phase the cells are placed in a specific location on the surface. Possibility of defects in metal and molecular QCA implementations are discussed in [54]. It is stated that wrong dot size or shape is possible because the targeted region is either under or over exposed during EBL. Single Electron Fault (SEF) in QCA are characterized and analysed in [55].

As mentioned in [56], occurrence of missing and extra dot or electrons is very less due to the ease of purification of small inorganic molecules. It is considered that these defects causes fatal errors and are easy to detect. Perhaps defects like cell misalignment, rotation, displacement, missing, addition occurs in the deposition phase are to be analyzed.

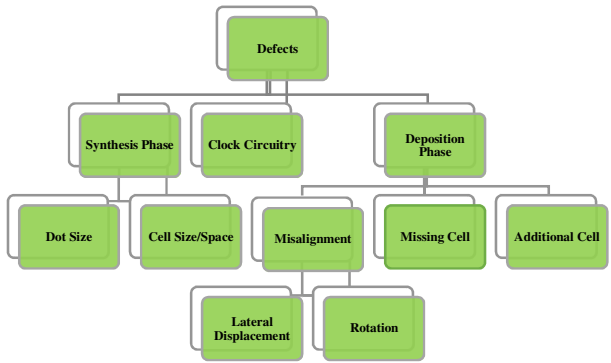

Fig. 9. QCA Defect Classification [48].

1. Cell Misalignment Defects

In misalignment defect, the cell of the defect free QCA device get displaced horizontally from its original position. If the cell is displace laterally then it is considered as cell displacement defects. Initially, Tahoori et al [57, 58] described the cell misalignment and displacement defects in basic QCA primitive MV, binary wire and inverter chain. Extensive simulation analysis of cell misalignment and displacement defects for MV is carried out using QCADesigner.

The displacement and misalignment of input cells of MV is shown Fig. 10 and 11 respectively [57]. Displacement defect in double inverter chain and binary wire is also reported in [57]. It observed that the horizontal cell input cell (input B) is most dominant cell. In MV, cell misalignment by the distance of greater or equal to half of the cell size causes undesired output. The cell displacement defect has less catastrophic effects on the functionality of an MV compared to cell misalignment defects. It is analysed that the double binary wire is more defect tolerant than the inverter chain in case of cell displacement defects.

Kink energy discussed earlier can be used to calculate the output cell polarization of any QCA device in case of cell misalignment and displacement defects.

The misalignment defects have more catastrophic effect on MV functionality compared to displacement defect, because if any cell get misaligned in right or left direction then the kink energy calculation differs and the output cell polarization may be undesired depending upon the distance. In cell displacement defect in MV, if cell get displaced by the distance less than or equal to the radius of effect then the output remains unaffected. Radius of effect is a distance from the centre of one cell to another cell which interacted with each other. Cells are not interacting if a cell to cell distance is more than the radius of effect. The radius of effect can be set during the simulation in QCADesigner tool.

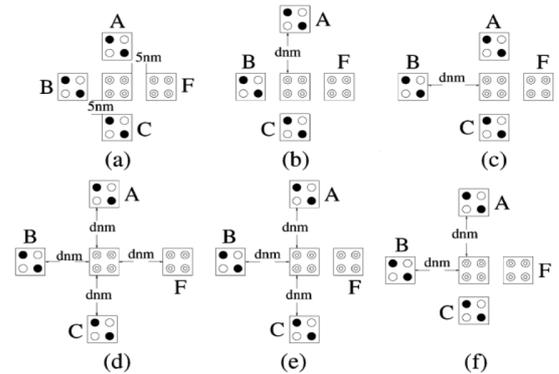

Fig. 10. Displacement in MV. (a) Fault free. (b) Displace A. (c) Displace B. (d) Displace all inputs and output. (e) Displace all inputs. (f) Displace A and B.

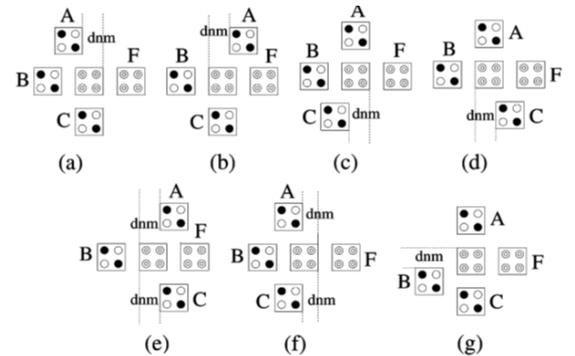

Fig. 11. Misalignments in MV (a) A misalignment (b) A misalignment (c) C misalignment (d) C misalignment (e) A, C misalignment (f) A, C misalignment (g) B misalignment.

*Yongqiang et al* [59] examined the effect of cell movement in horizontal and vertical directions at the same time (two-dimensions) for QCA fundamental devices MV, Inverter and wire. *Gabriel et al* [60] analyzed the behavior of QCA building blocks under the influence of random cell displacement defect. In [61], cell displacements of interconnect is analysed and maximum allowable displacement is determined.

2. Cell Rotation Defects

Yang et al [62] analysed the behaviour of QCA devices in presence of cell rotation. Further they have presented model for cell rotation effect using modified coherence vector formalism for permissible rotational angle.

3. Missing and Additional Cell Defects

During the lithography process, improper removal of resist causes extra cell attachment (additional) or missing cell defects. These defects also depends on the chemical compound used during the lithography. Modelling of additional cell can be done by adding extra cell to the periphery of device and circuit. In the same way modelling of missing cell can be done by removing the cell from device or circuit [56]. *Dysart et al* [56] analyzed the effects of missing cell defects in QCA wire assembled by a molecular implementation.

*Momenzadeh et al* [63] modeled single missing and additional cell defects in molecular QCA devices like Majority Voter (MV), inverter, fanout and L-Shaped. Also fault set corresponding to single missing and additional cell defects is proposed in [63] and mentioned in Table 3. It is observed that binary wire is less prone to the single missing cell defects. Additional cell does not alter the functionality of QCA devices except inverter for some cases. Huang et al [64] presented the single missing and additional cell defect characterization of sequential QCA circuits which is based on molecular QCA.

Table 3 Fault Set caused by single missing cell [63]

| Device | Fault Set |
|---|---|
| Majority Voter (MV) | S_a_B Maj (A', B, C') |
| Inverter | S_a_A |
| L-shaped wire | S_a_A' |
| Fanout | S_a_A' |

Simulation based single cell omission defect in MV, binary wire and wire crossing for 1-bit full adder has been reported in [57]. It is analyzed that the input cell B of MV is dominating cell. Output of MV doesn't altered due to missing of input cell B. Inversion of input A and C takes place in case of device missing cell. Single additional and missing cell defects in QCA sequential circuit are analyzed in [65].

Again the kink based energy calculation can be applied to calculate the polarization of device output cell in case of single missing and additional cell defect.

4. Defects in Clock Circuitry

Possibility of defect occurrence in clock circuitry is discussed in [66]. Phase shifts can result from manufacturing variations in each of the four required clock sources or from uneven path lengths [66]. The impact of QCA device scaling on defects is presented in [67]. Liu et al [68] explored the behavior of metal-dot QCA systems under stressed caused by the high temperature operation, high speed operation, and random variation in parameter values. Defect caused by fabrication variations in Magnetic Quantum-dot Cellular Automata at device, circuit, and architectural level is analyzed in [69]. Information-theoretic approach for investigation relationship between defect tolerance and redundancy in QCA devices is presented in [70].

Fault tolerant QCA devices and circuits are presented in [71-77]. Fault analysis of QCA combination circuit at layout and logic level is presented in [78]. Fabrication defects of a real molecular QCA wire built with ad hoc synthesized bis-ferrocene Molecules are analyzed in [79]. Also, in [80], the possible defects and causes of faults for a molecular QCA device are identified. The process variations effects in terms of yield and output error Rate for QCA based NanoMagnet Logic has been studied in [81].

*Apart from the available defect, multiple missing or additional cell defects can occur and possibly must be analyzed. Since self-assembly fabrication process for molecular QCA is prone to defects, in near future if QCA circuit and system exists, the key issue, defects must be addressed to avoid the failure of it. Looking into the reliability of QCA, extreme need for fault tolerant circuit is required. Also analysis of QCA oriented defects, its modeling and development of corresponding fault model is required.*

B. Faults, Fault models and Testing

The present-day technology uses single stuck at fault model (SSF) where the line is observed to be permanently stuck at either logic 0 (S_A_0) or logic 1 (S_A_1). The existing stuck-at-fault model can provide test generation for QCA circuits. Also, conventional ATPG tool can be exploited for QCA [58]

Initially the logic level testing of QCA is presented in [57]. Following properties for QCA MV with A, B, C as input lines of MV and Z as output line are investigated in [57-58, 82]

- For MV with input values *a*, *b* and *c* and output *z*, if all inputs are flipped, $abc \rightarrow a'b'c'$, then the output will also be flipped, i.e. $z \rightarrow z'$.
- If there is inversion at any input and/or the output of the MV, above property still holds
- The stuck-at-v fault on any input or output line of the voter is detectable by *abc* if and only if the stuck-at-*v'* fault on that line is detectable by *a'b'c'*
- If there are some inversions at any inputs and/or the output of the MV, then above property still holds
- Any vector that detects an input S_A_0 fault will also detect an output S_A_0 fault

The stuck at test set is developed to test all the simulated defects in MV [57]. It shown that the few test vector set is effectively detects all the simulated defects in MV with 100% test coverage. Design for Testability (DFT) scheme for QCA circuit is proposed in [57-58,82]. Two lines used to make MV as AND and OR gate can be viewed as control lines to implement DFT scheme in the network of MVs acting as AND and OR gates.

First comprehensive test generation methodology based on Boolean satisfiability (SAT) for combinational QCA circuits is proposed in [83,84]. The authors have shown that SSF set is not sufficient to detect all the simulated QCA defects in MV sothat further test generation is required for additional test vectors. Proposed ATPG generated test vectors are given to the fault simulator, the track on test vector applied to each MV is kept.

If there exists a majority gate which does not receive a 100% defect test set, additional test generation is performed. Once all the defects have been covered, a test set containing the SSF test set and additional QCA vectors is obtained. The proposed ATPG is tested on MCNC *(*Microelectronics Center of North Carolina*)* benchmark circuits. All of the benchmarks were first synthesized into multi-level majority networks using the logic synthesis tool MALS [48]. The bridging faults in QCA interconnects are also targeted in [83,84].

*Faisal Karim et al* [85] also developed the combinational ATPG using extended version of PODEM algorithm for majority and minority logic networks mostly targeting QCA. A genetic algorithm was used to fill-in the unspecified values in the test patterns produced by the ATPG in order to achieve compaction on the final test set size. The modified PODEM algorithm was tested on a set of MCNC benchmark circuits. Probability based controllability and observability approaches were taken into consideration to guide this ATPG.

*Probability based testability approach is applicable to the fanout-free combinational circuits. It fails for the circuits containing fanout points due to the reconvergent path. So, Sandia Controllability Observability Analysis Program (SCOAP) based testability approach can be the alternative to probability based testability approach.*

Basic fault model for single input missing cell deposition defect for QCA MV is developed in [86]. In [86], test generation for single input missing cell deposition defects in the QCA circuit consists of MV as AND and OR gate is carried out using the proposed properties and corresponding fault model. The proposed properties are as follows:

**Property 1**. If S_A_B fault is present at the output of MV as AND gate then assign A=0 or B' and B=1.

**Property 2.** If S_A_B fault is present at the output of MV as OR gate then assign A=1 or B' and B=0.

**Property 3.** While propagating the values at the output of MV as AND or OR gate, justify remaining fanin by the non-controlling value of the gate.

In [87], test vectors to exhaustively test the functionality of any 3-input majority gate and arithmatic circuit are generated to find crosstalk which is modelled as dominant bringing faults. Method for In-Circuit-Testing of QCA circuits is shown in [88].

*Available literature suggested the application of test generation method for current silicon technology to the QCA. Different type of test generation for QCA oriented fault models can be possible and explored. Also the conventional test generation algorithms and fault models can be explored for QCA combinational and sequential circuits.*

## IV. CONCLUSION AND DISCUSSION

The systematic survey on QCA, its defect analysis and testing method is carried out in this paper. Solid need of CAD tools for QCA circuit simulation and synthesis is required.

As occurrence of defects are possible in QCA devices due to the nanoscale nature, this paper more emphasizes on the QCA defect, fault model and testing.

Mostly the simulation based defect analysis is carried out in the referred literature. Defect analysis through mathematical representation is required. Apart from the available defects in synthesis and deposition phase, defects like multiple missing cells, multiple additional cell during the deposition phase must be looked upon.

Novel QCA oriented fault model must be evolved to develop the test generation methods. Perhaps, the available test generation algorithms can also be explored.

This paper will be useful to understand the QCA in depth for the beginners. Also, it will be helpful for the researcher to find the various area to work upon.

REFERENCES

[1] "International Technology Roadmap for Semiconductors (ITRS)", 2015 Edition, https://www.semiconductors.org/clientuploads/Research_Technology/ITRS/2015/6_2015 ITRS 2.0 Beyond CMOS.pdf
[2] Peercy, Paul S. "The drive to miniaturization." *Nature* 406, no. 6799 (2000): 1023-1026.
[3] Meindl, James D. "Beyond Moore's law: The interconnect era." *Computing in Science & Engineering* 5, no. 1 (2003): 20-24.
[4] Likharev, Konstantin K. "Single-electron devices and their applications." *Proceedings of the IEEE* 87, no. 4 (1999): 606-632.
[5] Lent, Craig S., and P. Douglas Tougaw. "Lines of interacting quantum-dot cells: A binary wire." *Journal of applied Physics* 74, no. 10 (1993): 6227-6233.
[6] Chen, Kevin J., Koichi Maezawa, and Masafumi Yamamoto. "InP-based high-performance monostable-bistable transition logic elements (MOBILEs) using integrated multiple-input resonant-tunneling devices." *IEEE Electron Device Letters* 17, no. 3 (1996): 127-129.
[7] Lent, Craig S., P. Douglas Tougaw, Wolfgang Porod, and Gary H. Bernstein. "Quantum cellular automata." *Nanotechnology* 4, no. 1 (1993): 49-57.
[8] Lent, Craig S., P. Douglas Tougaw, and Wolfgang Porod. "Quantum cellular automata: the physics of computing with arrays of quantum dot molecules." In *Physics and Computation, 1994. PhysComp'94, Proceedings. Workshop on*, pp. 5-13. IEEE, 1994.
[9] Lent, Craig S., and P. Douglas Tougaw. "A device architecture for computing with quantum dots." *Proceedings of the IEEE* 85, no. 4 (1997): 541-557.
[10] Orlov, A. O., I. Amlani, G. H. Bernstein, C. S. Lent, and G. L. Snider. "Realization of a functional cell for quantum-dot cellular automata." *Science* 277, no. 5328 (1997): 928-930.
[11] Amlani, Islamshah, Alexei O. Orlov, Gregory L. Snider, Craig S. Lent, and Gary H. Bernstein. "External charge state detection of a double-dot system." *Applied physics letters* 71, no. 12 (1997): 1730-1732.

[12] Lombardi, Fabrizo, and Jing Huang. Design and test of digital circuits by quantum-dot cellular automata. Artech House, Inc., 2007.
[13] Amlani, Islamshah, Alexei O. Orlov, Gregory L. Snider, Craig S. Lent, and Gary H. Bernstein. "Demonstration of a functional quantum-dot cellular automata cell." *Journal of Vacuum Science & Technology B: Microelectronics and Nanometer Structures Processing, Measurement, and Phenomena* 16, no. 6 (1998): 3795-3799.
[14] Amlani, Islamshah, Alexei O. Orlov, Gregory L. Snider, Craig S. Lent, and Gary H. Bernstein. "Demonstration of a six-dot quantum cellular automata system." *Applied Physics Letters* 72, no. 17 (1998): 2179-2181.
[15] Orlov, Alexei O., Ravi K. Kummamuru, Rajagopal Ramasubramaniam, Geza Toth, Craig S. Lent, Gary H. Bernstein, and Gregory L. Snider. "Experimental demonstration of a latch in clocked quantum-dot cellular automata." *Applied Physics Letters* 78, no. 11 (2001): 1625-1627.
[16] Lieberman, Marya, Sudha Chellamma, Bindhu Varughese, Yuliang Wang, Craig Lent, Gary H. Bernstein, Gregory Snider, and Frank C. Peiris. "Quantum-dot cellular automata at a molecular scale." *Annals of the New York Academy of Sciences* 960, no. 1 (2002): 225-239.
[17] Lent, Craig S., Beth Isaksen, and Marya Lieberman. "Molecular quantum-dot cellular automata." *Journal of the American Chemical Society* 125, no. 4 (2003): 1056-1063.
[18] Blair, Enrique P., and Craig S. Lent. "Quantum-dot cellular automata: an architecture for molecular computing." In *Simulation of Semiconductor Processes and Devices, 2003. SISPAD 2003. International Conference on*, pp. 14-18. IEEE, 2003.
[19] Lu, Yuhui, and Craig S. Lent. "Theoretical study of molecular quantum-dot cellular automata." *Journal of Computational Electronics* 4, no. 1-2 (2005): 115-118.
[20] Lent, Craig S., and Beth Isaksen. "Clocked molecular quantum-dot cellular automata." *IEEE Transactions on Electron Devices* 50, no. 9 (2003): 1890-1896.
[21] Cowburn, R. P., and M. E. Welland. "Room temperature magnetic quantum cellular automata." *Science* 287, no. 5457 (2000): 1466-1468.
[22] Bernstein, Gary H., Alexandra Imre, V. Metlushko, A. Orlov, L. Zhou, L. Ji, György Csaba, and Wolfgang Porod. "Magnetic QCA systems." *Microelectronics Journal* 36, no. 7 (2005): 619-624.
[23] Imre, Alexandra, G. Csaba, L. Ji, A. Orlov, G. H. Bernstein, and W. Porod. "Majority logic gate for magnetic quantum-dot cellular automata." *Science* 311, no. 5758 (2006): 205-208.
[24] Perez-Martinez, F., I. Farrer, D. Anderson, G. A. C. Jones, D. A. Ritchie, S. J. Chorley, and C. G. Smith. "Demonstration of a quantum cellular automata cell in a Ga As∕ Al Ga As heterostructure." *Applied physics letters* 91, no. 3 (2007): 032102.
[25] Lent, Craig S., and Gregory L. Snider. "The development of quantum-dot cellular automata." In *Field-Coupled Nanocomputing*, pp. 3-20. Springer Berlin Heidelberg, 2014.
[26] Toth, Geza, and Craig S. Lent. "Quasiadiabatic switching for metal-island quantum-dot cellular automata." *Journal of Applied physics* 85, no. 5 (1999): 2977-2984.
[27] Frost, Sarah E., Timothy J. Dysart, Peter M. Kogge, and C. S. Lent. "Carbon nanotubes for quantum-dot cellular automata clocking." In *Nanotechnology, 2004. 4th IEEE Conference on*, pp. 171-173. IEEE, 2004.
[28] Vankamamidi, Vamsi, Marco Ottavi, and Fabrizio Lombardi. "Two-dimensional schemes for clocking/timing of QCA circuits." *IEEE Transactions on Computer-Aided Design of Integrated Circuits and Systems* 27, no. 1 (2008): 34-44.
[29] Vankamamidi, Vamsi, Marco Ottavi, and Fabrizio Lombardi. "Clocking and cell placement for QCA." In *Nanotechnology, 2006. IEEE-NANO 2006. Sixth IEEE Conference on*, vol. 1, pp. 343-346. IEEE, 2006.
[30] Purohit, Prafull. *Ripple clock schemes for quantum-dot cellular automata circuits*. Rochester Institute of Technology, 2012.
[31] Lent, Craig S., Mo Liu, and Yuhui Lu. "Bennett clocking of quantum-dot cellular automata and the limits to binary logic scaling." *Nanotechnology* 17, no. 16 (2006): 4240.
[32] Tougaw, P. Douglas, and Craig S. Lent. "Logical devices implemented using quantum cellular automata." *Journal of Applied physics* 75, no. 3 (1994): 1818-1825.
[33] Parviz, B. Amir, Declan Ryan, and George M. Whitesides. "Using self-assembly for the fabrication of nano-scale electronic and photonic devices." *IEEE transactions on advanced packaging* 26, no. 3 (2003): 233-241.
[34] Hu, Wenchuang, Koshala Sarveswaran, Marya Lieberman, and Gary H. Bernstein. "High-resolution electron beam lithography and DNA nano-patterning for molecular QCA." *IEEE Transactions on Nanotechnology* 4, no. 3 (2005): 312-316.
[35] Niemier, Michael T., and Peter M. Kogge. "The" 4-diamond circuit"-a minimally complex nano-scale computational building block in qca." In *VLSI, 2004. Proceedings. IEEE Computer society Annual Symposium on*, pp. 3-10. IEEE, 2004.
[36] Jiao, J., G. L. Long, F. Grandjean, A. M. Beatty, and T. P. Fehiner. "Building Blocking for the Molecular Expressionof QCA, Isolation and Characterization of a Covalently Bounded Square Array of two Ferrocenium and TwoFerrocene Complexes." *Journal of the Am. Chem. Society (JACS Communications)* 125, no. 25 (2003): 7522-7523.
[37] Lu, Yuhui, and Craig Lent. "Self-doping of molecular quantum-dot cellular automata: mixed valence zwitterions." *Physical Chemistry Chemical Physics* 13, no. 33 (2011): 14928-14936.
[38] Blair, E. P., "Tools for the Design and Simulation of Clocked Molecular Quantum-dot Cellular Automata Circuits," Master's thesis, University of Notre Dame, Department of Electrical Engineering, 2003.
[39] Niemier, M., M. Kontz, and P. Kogge, "A design of and Design tools for A novel quantum dot based microprocessor*," in Proc. of the 37th Annual Design Automation Conference*, pp. 227–232, 2000.
[40] Teodósio, Tiago, and Leonel Sousa. "QCA-LG: A tool for the automatic layout generation of QCA combinational circuits." In *Norchip, 2007*, pp. 1-5. IEEE, 2007.
[41] K. Walus, T. Dysart, G. A. Jullien, and R. A. Budiman, "QCADesigner: A rapid design and simulation tool for quantum-dot cellular automata," *IEEE Trans. Nanotechnology,* vol. 3, no. 1, pp. 26–31, Mar. 2004.
[42] http://waluslab.ece.ubc.ca/qcadesigner/qca-designer-downloads/
[43] M. Ottavi, L. Schiano, and F. Lombardi, "HDLQ: A HDL environment for QCA design," *ACM J. Emerging Technol. Comput. Syst.,* vol. 2, no. 4, pp. 243–261, Oct. 2006.
[44] Tang, R., F. Zhang, and Y. B. Kim, "Quantum-Dot Automata SPICE Macro Model", *ACM Great Lake Symposium on VLSI 2005*, 2005, pp. 108-111.
[45] Srivastava, S., et al., "QCAPro-An Error-Power Estimation Tool for QCA Circuit Design*," in Proc. of the IEEE International Symposium on Circuits and Systems,* pp. 2377–2380, 2011.
[46] Ottavi, Marco, et al. "An HDL Model of Magnetic Quantum-Dot Cellular Automata Devices and Circuits." Nanoelectronic Device Applications Handbook
[47] Zhang, R., et al., "A method of majority logic reduction for quantum cellular automata," *IEEE Transactions on Nanotechnology,* vol 3 (4), pp. 443-450, 2004.
[48] Zhang, R., P. Gupta, N. K. Jha, "Synthesis of majority and minority networks and its application to QCA, TPL, and SET based nanotechnologies", *IEEE Conference on VLSI Design held jointly with International Conference on Embedded Systems Design,* 2005, pp. 229-234.
[49] Kong, Kun, Yun Shang, and Ruqian Lu. "An optimized majority logic synthesis methodology for quantum-dot cellular automata." *IEEE Transactions on Nanotechnology* 9, no. 2 (2010): 170-183.
[50] Wang, Peng, Mohammed Y. Niamat, Srinivasa R. Vemuru, Mansoor Alam, and Taylor Killian. "Synthesis of Majority/Minority Logic Networks." *IEEE Transactions on Nanotechnology* 14, no. 3 (2015): 473-483.
[51] Devadoss, Rajeswari, Kolin Paul, and M. Balakrishnan. "MajSynth: Ann-inputMajorityAlgebrabasedLogicSynthesisToolfor Quantum-dotCellularAutomata." (2015).
[52] Mishra, Vipul Kumar, and Himanshu Thapliyal. "Heuristic Based Majority/Minority Logic Synthesis for Emerging Technologies." In *VLSI Design and 2017 16th International Conference on Embedded Systems (VLSID), 2017 30th International Conference on*, pp. 295-300. IEEE, 2017.
[53] Dhare, Vaishali, and Usha Mehta. "Defect characterization and testing of QCA devices and circuits: A survey." In *VLSI Design and Test (VDAT), 2015 19th International Symposium on*, pp. 1-2. IEEE, 2015.

[54] Dysart, Timothy J., and Peter M. Kogge. "Strategy and prototype tool for doing fault modeling in a nano-technology." In *Nanotechnology, 2003. IEEE-NANO 2003. 2003 Third IEEE Conference on*, vol. 1, pp. 356-359. IEEE, 2003.

[55] Mukherjee, Rijoy, et al. "Characterization and analysis of single electron fault of QCA primitives." *Microelectronics, Computing and Communications (MicroCom), 2016 International Conference on*. IEEE, 2016.

[56] Dysart, Timothy J., Peter M. Kogge, Craig S. Lent, and Mo Liu. "An analysis of missing cell defects in quantum-dot cellular automata." In *IEEE International Workshop on Design and Test of Defect-Tolerant Nanoscale Architectures (NANOARCH)*, vol. 3, pp. 1-8. 2005.

[57] Tahoori, Mehdi Baradaran, Mariam Momenzadeh, Jin Huang, and Fabrizio Lombardi. "Defects and faults in quantum cellular automata at nano scale." In *VLSI Test Symposium, 2004. Proceedings. 22nd IEEE*, pp. 291-296. IEEE, 2004.

[58] Tahoori, Mehdi B., Jing Huang, Mariam Momenzadeh, and Fabrizio Lombardi. "Testing of quantum cellular automata." *IEEE Transactions on Nanotechnology* 3, no. 4 (2004): 432-442.

[59] Zhang, Yongqiang, Hongjun Lv, Shuai Liu, Yunlong Xiang, and Guangjun Xie. "Defect-tolerance analysis of fundamental quantum-dot cellular automata devices." *The Journal of Engineering* 1, no. 1 (2015).

[60] Schulhof, Gabriel, Konrad Walus, and Graham A. Jullien. "Simulation of random cell displacements in QCA." *ACM Journal on Emerging Technologies in Computing Systems (JETC)* 3, no. 1 (2007): 2.

[61] Karim, Faizal, and Konrad Walus. "Characterization of the displacement tolerance of QCA interconnects." *Design and Test of Nano Devices, Circuits and Systems, 2008 IEEE International Workshop on*. IEEE, 2008.

[62] Yang, Xiaokuo, Li Cai, Shuzhao Wang, Zhuo Wang, and Chaowen Feng. "Reliability and performance evaluation of QCA devices with rotation cell defect." *IEEE Transactions on Nanotechnology* 11, no. 5 (2012): 1009-1018.

[63] Momenzadeh, Mariam, Marco Ottavi, and Fabrizio Lombardi. "Modeling QCA defects at molecular-level in combinational circuits." In *Defect and Fault Tolerance in VLSI Systems, 2005. DFT 2005. 20th IEEE International Symposium on*, pp. 208-216. IEEE, 2005.

[64] Huang, Jing, Mariam Momenzadeh, and Fabrizio Lombardi. "Analysis of missing and additional cell defects in sequential quantum-dot cellular automata." *INTEGRATION, the VLSI journal* 40, no. 4 (2007): 503-515.

[65] Momenzadeh, Mariam, Jing Huang, and Fabrizio Lombardi. "Defect characterization and tolerance of QCA sequential devices and circuits." In *Defect and Fault Tolerance in VLSI Systems, 2005. DFT 2005. 20th IEEE International Symposium on*, pp. 199-207. IEEE, 2005.

[66] Ottavi, Marco, Hamid Hashempour, Vamsi Vankamamidi, Faizal Karim, Konrad Walus, and André Ivanov. "On the error effects of random clock shifts in quantum-dot cellular automata circuits." In *Defect and Fault-Tolerance in VLSI Systems, 2007. DFT'07. 22nd IEEE International Symposium on*, pp. 487-498. IEEE, 2007.

[67] Huang, Jing, Mariam Momenzadeh, Mehdi Baradaran Tahoori, and Fabrizio Lombardi. "Defect characterization for scaling of QCA devices [quantum dot cellular automata]." In *Defect and Fault Tolerance in VLSI Systems, 2004. DFT 2004. Proceedings. 19th IEEE International Symposium on*, pp. 30-38. IEEE, 2004.

[68] Liu, Mo, and Craig S. Lent. "Reliability and defect tolerance in metallic quantum-dot cellular automata." *Journal of Electronic Testing* 23, no. 2-3 (2007): 211-218.

[69] Niemier, Michael, Michael Crocker, and X. Sharon Hu. "Fabrication variations and defect tolerance for nanomagnet-based QCA." In *Defect and Fault Tolerance of VLSI Systems, 2008. DFTVS'08. IEEE International Symposium on*, pp. 534-542. IEEE, 2008.

[70] Dai, Jianwei, Lei Wang, and Fabrizio Lombardi. "An information-theoretic analysis of quantum-dot cellular automata for defect tolerance." *ACM Journal on Emerging Technologies in Computing Systems (JETC)* 6, no. 3 (2010): 9.

[71] Wei, Tongquan, Kaijie Wu, Ramesh Karri, and Alex Orailoglu. "Fault tolerant quantum cellular array (QCA) design using triple modular redundancy with shifted operands." In *Proceedings of the 2005 Asia and South Pacific Design Automation Conference*, pp. 1192-1195. ACM, 2005.

[72] Ma, Xiaojun, and Fabrizio Lombardi. "Fault tolerant schemes for QCA systems." In *Defect and Fault Tolerance of VLSI Systems, 2008. DFTVS'08. IEEE International Symposium on*, pp. 236-244. IEEE, 2008.

[73] Dalui, Mamata, Bibhash Sen, and Biplab K. Sikdar. "Fault tolerant QCA logic design with coupled majority-minority gate." *Int. J. Comput. Appl* 1, no. 29 (2010): 81-87.

[74] Farazkish, Razieh. "A new quantum-dot cellular automata fault-tolerant five-input majority gate." *Journal of nanoparticle research* 16, no. 2 (2014): 2259.

[75] Roohi, Arman, Ronald F. DeMara, and Navid Khoshavi. "Design and evaluation of an ultra-area-efficient fault-tolerant QCA full adder." *Microelectronics Journal* 46, no. 6 (2015): 531-542.

[76] Farazkish, Razieh. "A new quantum-dot cellular automata fault-tolerant full-adder." *Journal of Computational Electronics* 14, no. 2 (2015): 506-514.

[77] Sen, Bibhash, Yashraj Sahu, Rijoy Mukherjee, Rajdeep Kumar Nath, and Biplab K. Sikdar. "On the reliability of majority logic structure in quantum-dot cellular automata." *Microelectronics Journal* 47 (2016): 7-18.

[78] Dhare, Vaishali, and Usha Mehta. "Fault analysis of QCA combinational circuit at layout & logic level." In *Electrical and Computer Engineering (WIECON-ECE), 2015 IEEE International WIE Conference on*, pp. 22-26. IEEE, 2015.

[79] Pulimeno, Azzurra, Mariagrazia Graziano, Alessandro Sanginario, Valentina Cauda, Danilo Demarchi, and Gianluca Piccinini. "Bis-ferrocene molecular QCA wire: Ab initio simulations of fabrication driven fault tolerance." *IEEE transactions on nanotechnology* 12, no. 4 (2013): 498-507.

[80] Graziano, Mariagrazia, Azzurra Pulimeno, Ruiyu Wang, Xiang Wei, Massimo Ruo Roch, and Gianluca Piccinini. "Process variability and electrostatic analysis of molecular QCA." *ACM Journal on Emerging Technologies in Computing Systems (JETC)* 12, no. 2 (2015): 18.

[81] Turvani, Giovanna, Fabrizio Riente, Mariagrazia Graziano, and Maurizio Zamboni. "A quantitative approach to testing in quantum dot cellular automata: Nanomagnet logic case." In *Ph. D. Research in Microelectronics and Electronics (PRIME), 2014 10th Conference on*, pp. 1-4. IEEE, 2014.

[82] Tahoori, Mehdi Baradaran, and Fabrizio Lombardi. "Testing of quantum dot cellular automata based designs." In *Proceedings of the conference on Design, automation and test in Europe-Volume 2*, p. 21408. IEEE Computer Society, 2004.

[83] Gupta, Pallav, Niraj K. Jha, and Loganathan Lingappan. "A test generation framework for quantum cellular automata circuits." *IEEE transactions on very large scale integration (VLSI) systems* 15, no. 1 (2007): 24-36.

[84] Gupta, Pallav, Niraj K. Jha, and Loganathan Lingappan. "Test generation for combinational quantum cellular automata (QCA) circuits." In *Design, Automation and Test in Europe, 2006. DATE'06. Proceedings*, vol. 1, pp. 1-6. IEEE, 2006.

[85] Karim, Faizal, Konrad Walus, and Andre Ivanov. "Testing of combinational majority and minority logic networks." In *Mixed-Signals, Sensors, and Systems Test Workshop, 2008. IMS3TW 2008. IEEE 14th International*, pp. 1-6. IEEE, 2008.

[86] Dhare, Vaishali, and Usha Mehta. "Development of basic fault model and corresponding ATPG for single input missing cell deposition defects in Majority Voter of QCA." In *Region 10 Conference (TENCON), 2016 IEEE*, pp. 2354-2359. IEEE, 2016.

[87] Karim, Faizal, Konrad Walus, and André Ivanov. "Crosstalk in QCA arithmetic circuits." In *PROCEEDINGS-SPIE THE INTERNATIONAL SOCIETY FOR OPTICAL ENGINEERING*, vol. 6313, p. 631306. International Society for Optical Engineering; 1999, 2006.

[88] Kazemi-fard, Nasim, Maryam Ebrahimpour, Mostafa Rahimi, Mohammad Tehrani, and Keivan Navi. "Performance evaluation of in-circuit testing on QCA based circuits." In *Design & Test Symposium (EWDTS), 2008 East-West*, pp. 375-378. IEEE, 2008.